\documentclass{emulateapj}
\newcommand{\gps}{\ensuremath{g_{\rm P1}}}
\newcommand{\rps}{\ensuremath{r_{\rm P1}}}
\newcommand{\ips}{\ensuremath{i_{\rm P1}}}
\newcommand{\zps}{\ensuremath{z_{\rm P1}}}
\newcommand{\yps}{\ensuremath{y_{\rm P1}}}

\newcommand{\PS}{\protect \hbox {Pan-STARRS1}}
\newcommand{\degree}{\ensuremath{^\circ}}
\newcommand{\dfplot}[1]{\plotone{figs/#1}}
\newcommand{\minst}{{\ensuremath{m_{\mathrm{inst}}}}}

\shorttitle{Preliminary PS1 Photometric Calibration}
\shortauthors{E. F. Schlafly et al.}
\begin{document}
\title{Photometric Calibration of the First 1.5 Years of the Pan-STARRS1 Survey}
%
%
%
\author{
E. F. Schlafly,\altaffilmark{1}
D. P. Finkbeiner,\altaffilmark{1,2}
M. Juri\'c,\altaffilmark{2,3}
E. A. Magnier,\altaffilmark{4}
W. S. Burgett,\altaffilmark{4}
K. C. Chambers,\altaffilmark{4} 
T. Grav,\altaffilmark{5}
K. W. Hodapp,\altaffilmark{4}
N. Kaiser,\altaffilmark{4}
R.-P. Kudritzki,\altaffilmark{4}
N. F. Martin,\altaffilmark{6,7}
J. S. Morgan,\altaffilmark{4}
P. A. Price,\altaffilmark{8}
H.-W. Rix,\altaffilmark{6}
C. W. Stubbs,\altaffilmark{1,2}
J. L. Tonry\altaffilmark{4} and
R. J. Wainscoat\altaffilmark{4}
}

\altaffiltext{1}{Department of Physics, Harvard University, 17 Oxford Street, Cambridge MA 02138}
\altaffiltext{2}{Harvard-Smithsonian Center for Astrophysics, 60 Garden Street, Cambridge, MA 02138}
\altaffiltext{3}{Hubble Fellow}
\altaffiltext{4}{Institute for Astronomy, University of Hawaii, 2680 Woodlawn Drive, Honolulu HI 96822}
\altaffiltext{5}{Planetary Science Institute, 1700 East Fort Lowell, Suite 106, Tucson, AZ 8579, USA}
\altaffiltext{6}{Max Planck Institute for Astronomy, K\"{o}nigstuhl 17, D-69117 Heidelberg, Germany}
\altaffiltext{7}{Observatoire Astronomique de Strasbourg, CNRS, UMR 7550, 11 rue de l'Universit\'{e}, F-67000 Strasbourg, France}
\altaffiltext{8}{Department of Astrophysical Sciences, Princeton University, Princeton, NJ 08544, USA}

%
%
\begin{abstract}
We present a precise photometric calibration of the first 1.5 years of science imaging from the Pan-STARRS1 survey (PS1), an ongoing optical survey of the entire sky north of declination $-30\degree$ in five bands.  Building on the techniques employed by \citet{Padmanabhan:2008} in the Sloan Digital Sky Survey (SDSS), we use repeat PS1 observations of stars to perform the relative calibration of PS1 in each of its five bands, solving simultaneously for the system throughput, the atmospheric transparency, and the large-scale detector flat field.  Both internal consistency tests and comparison against the SDSS indicate that we achieve relative precision of $<10$ mmag in $g$, $r$, and $i_{\rm P1}$, and $\sim 10$ mmag in $z$ and $y_{\rm P1}$.  The spatial structure of the differences with the SDSS indicates that errors in both the PS1 and SDSS photometric calibration contribute similarly to the differences.  The analysis suggests that both the PS1 system and the Haleakala site will enable $<1$\%  photometry over much of the sky.
\end{abstract}



\section{Introduction}
\label{sec:intro}

A central problem in astronomy is relating the number of photons recorded at a detector to the physical flux of photons emitted from a source.  This relation depends on important astrophysical parameters, like the distance and Galactic extinction to the source, as well as more ephemeral, local phenomena like the weather at the telescope and the sensitivity of the detector.  The problem of photometric calibration is to characterize these latter phenomena to render more universal the detected astronomical phenomena.

The photometric calibration of optical data is often performed by comparing multiple observations of the same sources for large sets of sources and demanding that their fluxes not change over time.  This is the same technique used to calibrate cosmic microwave background and radio data, and it has been extensively used in optical astronomy, e.g. \citet{Maddox:1990}, \citet{Honeycutt:1992}, \citet{Fong:1992, Fong:1994}, \citet{Glazebrook:1994}, and \citet{Magnier:1992}.  In the Sloan Digital Sky Survey \citet[SDSS]{York:2000}, \citet{Padmanabhan:2008} applied this technique and achieved a photometric calibration accurate at the 1\% level.

Often surveys and observations have been calibrated using repeat observations of a small number of fields of standard stars, e.g., the 2 Micron All Sky Survey \citep{Skrutskie:2006}.  These calibration observations determine the relation between flux and photon-count when the fields are observed, and the relation during science observations is extrapolated from them.  In the SDSS, however, each set of observations was made to slightly overlap other observations, and the network of these overlaps was used to simultaneously calibrate all of the SDSS observations.  Upcoming optical wide-area surveys like PS1, DES, and LSST plan to improve upon this technique by tremendously increasing the number of multiply-observed stars; each plans to image their entire survey area several times.  The dense overlapping regions of these upcoming surveys should yield a much more tightly constrained photometric calibration.

Current surveys require photometric calibration as accurate as possible, ideally to better than the percent level.  Typical photometric uncertainties from point-spread-function modeling reach the 1\% level, and absent an equally good photometric calibration, calibration errors will dominate this uncertainty.  Additionally, the width of the stellar locus is about 1\% in certain color combinations \citep{Ivezic:2007}, requiring an equally good calibration to allow stars to be photometrically identified most accurately.  Studies of the interstellar dust at high Galactic latitudes can be still more demanding; photometric calibration dominates the uncertainty in the analysis of \citet{Schlafly:2011}.  Uncertainty in the photometric calibration even at the 1\% level can contribute to significant variation in the number densities of galaxies used for clustering studies on large angular scales \citep{Ross:2012}.  In short, a number of current science projects are limited by photometric calibration accuracy.

In this work, we describe our application of the \citet{Padmanabhan:2008} photometric calibration algorithm to the first 1.5 years of PS1 survey data to achieve a 1\% calibration.  This paper is organized as follows: in Section \textsection \ref{sec:ps1}, we describe the PS1 survey and its current status.  Section \textsection \ref{sec:methods} describes the photometric calibration algorithm and its application to PS1 data.  Section \textsection \ref{sec:results} gives the results of our calibration of the PS1 data, and the results of the tests used to verify the calibration.  In \textsection \ref{sec:discussion}, we discuss the stability of the PS1 system and the atmosphere in light of these results.  Finally, in Section \textsection \ref{sec:conclusion} we summarize, mention prospects for the future, and conclude.

\subsection{The Pan-STARRS1 System and Surveys}
\label{sec:ps1}

The Pan-STARRS1 system is a wide-field optical imager devoted to survey operations \citep{PS1_system}.  The telescope has a 1.8~meter diameter primary mirror, located on the peak of Haleakala on Maui \citep{PS1_optics}.  The site and optics deliver a point spread function with a full-width at half-maximum (FWHM) of about one arcsecond, over a seven square degree field of view.  The focal plane of the telescope is equipped with the Gigapixel Camera 1 (GPC1), an array of 60 $4800\times4800$ orthogonal transfer array (OTA) CCDs \citep{PS1_GPCA, PS1_GPCB}.  Each OTA CCD is further subdivided into an $8\times8$ array of independently-addressable detector regions which are individually read out by the camera electronics through their own on-chip amplifier.

Most of the PS1 observing time is dedicated to two surveys: the $3\pi$ survey, a survey of the entire sky north of declination $-30\degree$, and the medium-deep (MD) survey, a deeper, many-epoch survey of 10 fields, each 7 $\rm{deg}^2$ in size \citep{PS_MDRM}.  Each survey is conducted in five broadband filters, denoted \gps, \rps, \ips, \zps, and \yps, that together span 400--1000 nm.  These filters are similar to those used in the SDSS, except the \gps\ filter extends 20~nm redward of $g_{\rm SDSS}$ while the \zps\ filter is cut off at 920 nm.  The \yps\ filter covers the region from 920nm to 1030nm with the red limit largely determined by the transparency of the silicon in the detector.  These filters and their absolute calibration in the context of \PS\ are described in \citet{PS_lasercal} and \citet{JTphoto}.  The filter bandpasses are summarized in Table \ref{tab:filt}.

\begin{deluxetable}{cccccc}
\tablewidth{\columnwidth}
\tablecaption{Pan-STARRS1 Bandbass Parameters}
\tablehead{
\colhead{Filter} & \colhead{$\lambda_{\mathrm{eff}}$} & \colhead{$\lambda_{\mathrm{B}}$} & \colhead{$\lambda_{\mathrm{R}}$} & \colhead{ZP} & \colhead{$\mu$}
}
\startdata
\gps & 481 & 414 & 551 & 24.41 & 21.92 \\
\rps & 617 & 550 & 689 & 24.68 & 20.83 \\
\ips & 752 & 690 & 819 & 24.56 & 19.79 \\
\zps & 866 & 818 & 922 & 24.22 & 19.24 \\
\yps & 962 & 918 & 1001 & 23.24 & 18.24
\enddata
\tablecomments{
\label{tab:filt}
\PS\ bandpass parameters.  The column $\lambda_{\mathrm{eff}}$ gives the effective wavelength of each filter in nm, while the columns $\lambda_{\mathrm{B}}$ and $\lambda_{\mathrm{R}}$ give the filter blue and red cutoffs in nm.  The AB zero points are given by ZP, and observed sky brightness in magnitudes per square arcsecond by $\mu$.  All values are from \citet{JTphoto}, except for the zero points.  These are marginally discrepant from the values of \citet{JTphoto} due to the variation in throughput over the PS1 focal plane, and are only intended to serve as a rough guide.
}
\end{deluxetable}

The PS1 images are processed by the \PS\ Image Processing Pipeline (IPP) \citep{PS1_IPP}.  This pipeline performs automatic bias subtraction, flat fielding, astrometry \citep{PS1_astrometry}, photometry \citep{PS1_photometry}, and image stacking and differencing for every image taken by the system.  The approximately one trillion pixels per night are processed in a massively parallel fashion at the Maui High Performance Computer Center.

The $3\pi$ survey is executed so that each time a patch of sky is visited, it is observed for about 40 seconds twice, at times separated by an interval of about 15 minutes \citep{PS_MDRM}.  The two observations make a transit-time-interval (TTI) pair.  These observations are used primarily to search for high proper-motion Solar System objects.  Each year, the field is then observed a second time in the same filter with an additional TTI pair of images, making for four images of each part of the sky, per year, in each of the 5 PS1 filters.  The MD observations consist of 8, much longer, $\sim 200$ second exposures, dithered in both position and position angle.

\begin{figure*}[tbh]
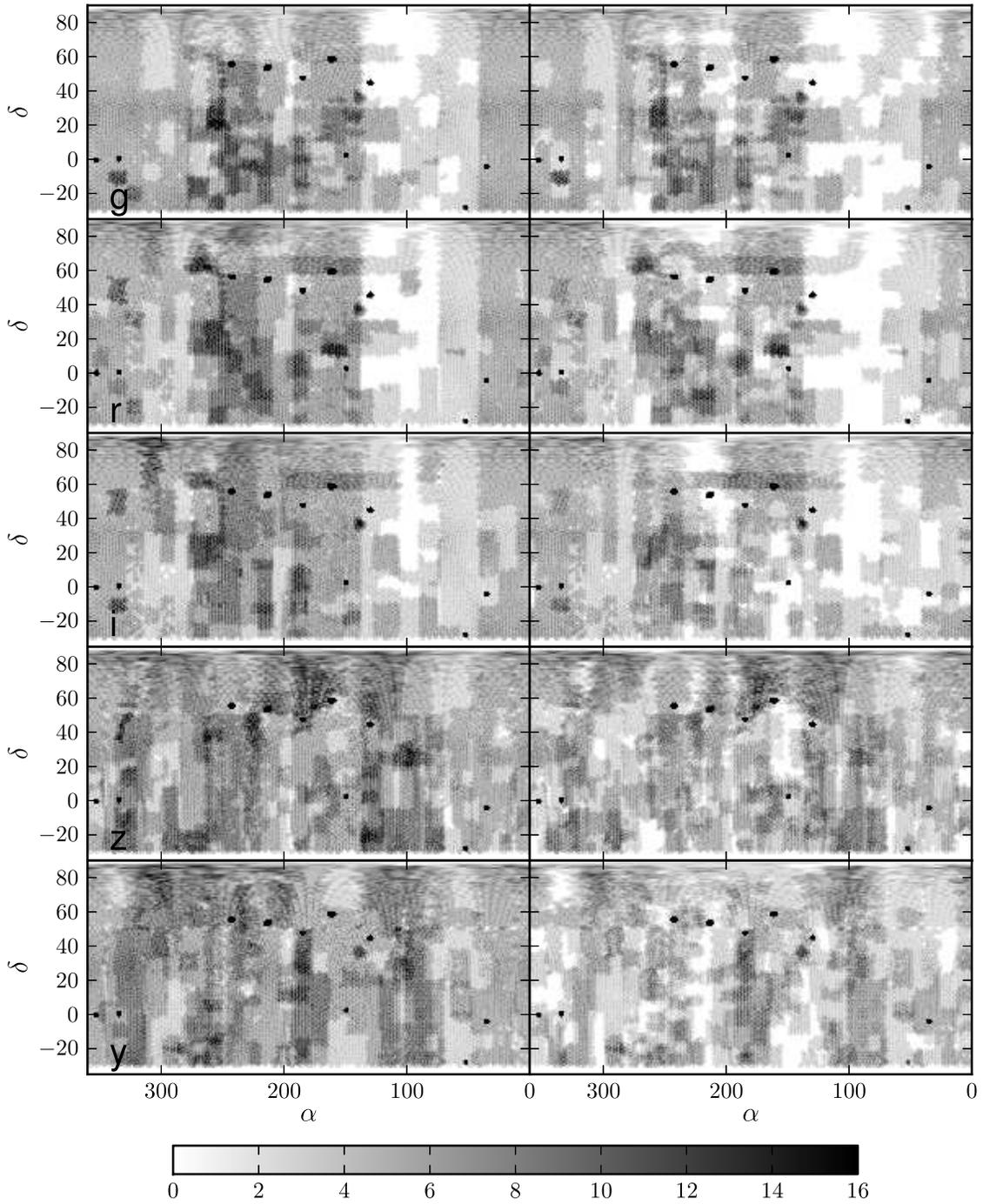

\dfplot{coverage.eps}
\figcaption{
\label{fig:coverage}
Map of the number of times the sky has been observed by \PS, overall (left panels) and in photometric weather (right panels) (see \textsection \ref{subsec:algops1}).  The x-axes give right ascension and the y-axes give declination, both in degrees.  White to black spans 0 to 16.  The ten black circles are the locations of the 10 medium-deep fields, which are observed more frequently than the rest of the survey area.  In the winter of 2010, the combination of bad weather and the malfunction of the \PS\ shutter suspended operations, leading to an area of poor coverage in $gri$ around right ascension 100\degree.
}
\end{figure*}

The data used in this analysis were taken primarily between February 12, 2010 and June 19, 2011, though a small amount of data from as early as June 20, 2009 is included.  Figure \ref{fig:coverage} shows the number of times each part of the sky was observed during this period.  The left panel gives the total number of times the sky was observed, while the right panel gives the number of times the sky was observed in photometric conditions (\textsection \ref{subsec:algops1}).  The median number of times each part of the sky was observed in photometric weather is 4 in each band. This makes for two independent TTI pairs of observations, on average, of each part of the sky, though Figure \ref{fig:coverage} makes clear that this coverage is variable, and the sky around right ascension 100\degree\ is covered only by one TTI pair of observations or not at all in $gr\ips$. The MD fields have been observed much more often; 100--300 times in $gri\zps$, and about 100 times in \yps.

\section{Methods}
\label{sec:methods}

Like all CCD-equipped telescopes, PS1 ultimately records the number of photons received from objects it targets\footnote{The gain of the GPC1 camera is nearly 1 ADU/electron, and we include the quantum efficiency (electrons per photon) in the detector throughput, so we do not distinguish between ADUs and photons in this discussion}.  The number of photons recorded depends on:
\begin{enumerate}
\item the transparency of the night sky toward the object,
\item the throughput of the detector, filter, and optics, and
\item the size and reflectivity of the telescope mirror.
\end{enumerate}
The object of the photometric calibration is to convert the measurements of the number of photons recorded by the system to measurements of the incident flux from the object, eliminating the signatures of the instrument and atmosphere.  The calibration can be divided into two separate procedures, as described in \citet{Padmanabhan:2008}: the relative photometric calibration, in which the differences in system throughput from observation to observation are removed, and the absolute photometric calibration, in which the number of photons recorded for some particular configuration of the telescope is converted to a magnitude on the AB magnitude system, which is based on physical units of flux \citep{Oke:1983}.  This paper presents the relative calibration of the \PS\ system; the absolute calibration is described in \citet{JTphoto}.  Given an absolute flux calibration for a single star, the relative calibration transfers this absolute calibration over every observation of the survey.

In this section, we describe the method used to perform the relative calibration of the \PS\ data.  First, in \textsection \ref{subsec:problem}, we describe the general problem of the photometric calibration of optical data.  In \textsection \ref{subsec:algorithm}, we describe the algorithm used to perform the relative calibration of the survey, that of \citet{Padmanabhan:2008}.  In \textsection \ref{subsec:algops1}, we then describe the details of the implementation of the algorithm for processing PS1 data.

\subsection{The Goal of Photometric Calibration}
\label{subsec:problem}

The quantities of scientific interest in an imaging survey are usually the astrometry and photometry of objects as a function of time.  The photometry of an object gives the flux from that object reaching the earth within a filter bandpass.  For linear detectors like the PS1 CCDs, instead the number of photons per second $N$ reaching the detectors is directly measured, which is simply related to the flux $f$ by 
\begin{equation}
\label{eq:fluxcal}
N = Kf
\end{equation}
if noise is ignored.  The task of the photometric calibration is to solve for the throughput $K$.

Conventionally, the photometry of an object is given in magnitudes $m$, with $m = -2.5\log{f/f_0}$, where $f_0$ gives the AB magnitude reference flux \citep{Oke:1983}.  Then Equation \ref{eq:fluxcal} becomes $\minst + Z = m$, with the instrumental magnitude given by $\minst = -2.5\log{N}$ and the zeropoint given by $Z = 2.5\log{Kf_0}$.  We work in these logarithmic variables for the rest of this work, and so seek to determine the zero points $Z$ of each observation in the survey.

The zero points $Z$ are determined by the light collecting efficiency of all of the components of the system.  At a particular wavelength $\lambda$, $Z$ can be decomposed into a number of factors describing the system: 
\begin{equation}
Z = 2.5\log{AT_aT_oT_fT_df_0}
\end{equation}
Here $A$ gives the collecting area of the telescope, and $T_a$, $T_o$, $T_f$, and $T_d$ the throughputs of the atmosphere, optics, filter, and detector, respectively.  In principle, $Z$ can be different for each star in each exposure, if, for instance, the filter throughput or detector efficiency varies over the focal plane.

The relative calibration of the survey is concerned only with how $Z$ varies from object to object in the survey.  We can then separate $Z$ into two terms, $Z = Z_a + Z_r$, where $Z_a$ is a constant giving the absolute zero point of the survey in a particular situation, and $Z_r$ gives the relative change in zero point from $Z_a$.  Unchanging components of $Z$, like the reference flux $f_0$ and mirror area $A$, affect only $Z_a$ and are independent of $Z_r$.  The model for $Z_r$ must be simple enough to allow its components to be constrained, but flexible enough to capture the variation in $Z_r$.

For a system like PS1, where the optical system, filter, and detector are essentially unchanged over the course of the night, we can suppose that $A$, $T_o$, $T_f$, and $T_d$ are constant over the course of the night.  We can then encapsulate the effect of all these terms on the zero point as a single term, $a$, for that night, with $a = 2.5\log{AT_oT_fT_d}$, and seek to measure how $a$ varies over the course of the survey.  

We must also account for $T_a$, the variation in the transparency of the atmosphere.  We model the atmospheric transparency simply as 
\begin{equation}
\label{eq:airmass}
2.5\log{T_a} = -k x
\end{equation}
where $k$ describes the effectiveness of the atmosphere at extinguishing light, and $x$ is the airmass of the observation \citep{Padmanabhan:2008}.  The survey is executed so that all observations have low airmass; the largest airmass included is 2.7 and the vast majority of images have airmass less than 1.6.  Equation \ref{eq:airmass} strictly holds only for monochromatic light, or when the atmospheric extinction does not vary with wavelength.  This assumption is violated in the \yps\ band, where there are strong water absorption features in the atmospheric extinction (see \textsection \ref{sec:discussion} and \citet{JTphoto}).  Moreover, $k$ will only be constant over a night if the atmosphere is isotropic and unchanging in time, an assumption clearly violated when clouds are present.  Still, we find that for most of the nights of the survey, the simple model $2.5\log{T_a} = -kx$ is largely satisfactory (but see \textsection \ref{subsec:systemstability} for more details).

We therefore present as a starting point for our photometric model the simple expression $Z = a - kx$, consistent with the above discussion.  We ultimately adopt a more complicated variation of this function in \textsection \ref{subsec:algops1}.  The problem of photometric calibration then becomes to determine the parameters $a$ and $k$ of such a model for each night of the survey.  We perform this calibration following the algorithm of \citet{Padmanabhan:2008}, finding the parameters that minimize the variance of repeat observations of each star.

\subsection{Algorithm}
\label{subsec:algorithm}

An optical survey provides instrumental magnitudes \minst\ of objects in the sky.  We may have several repeated observations of the same object, $m_{\mathrm{inst},o,i}$ with uncertainties $\sigma_{o,i}$, where $o$ labels an object and $i$ labels its observations.  We are ultimately interested in the calibrated magnitude $m$ of the object, with $m = \minst + Z_a + Z_r$, where $Z_a$ and $Z_r$ are defined in \textsection \ref{subsec:problem}.  We find $Z_r$ by minimizing
\begin{equation}
\sum_o \sum_i (m_{o,i} - \overline{m_o})^2/\sigma_{o,i}^2
\end{equation}
where $\overline{m}_o$ is the average of $m_{o,i}$ over all observations of object $o$.  The absolute zero point $Z_a$ cancels out of this expression.  

Letting $\mathbf{m}$ be a vector with every observation of every object in the survey, and using a linear model for $Z$, then $\mathbf{m} = \mathbf{\minst} + \mathbf{A} \mathbf{p}$, where $\mathbf{A}$ is the design matrix for $Z$ and $\mathbf{p}$ contains the parameters of the model for $Z$.  For the simple model for $Z$ described in \textsection \ref{subsec:problem}, $\mathbf{p}$ is a vector containing a parameter $a$ and $k$ for each night of the survey.  The vector $\mathbf{m}$ has length $n_{\mathrm{obs}}$, the total number of observations of all objects in the survey.  The matrix $\mathrm{A}$ has dimensions $n_\mathrm{obs}\times n_\mathrm{par}$, where $n_\mathrm{par}$ is the number of parameters in the model.  Furthermore, let $\mathbf{W}$ be the $n_\mathrm{obs}\times n_\mathrm{obs}$ matrix of weights, such that if $\mathbf{m}_j$ corresponds to an observation of object $o$, then
\begin{equation}
\sum_i W_{i,j}m_i = \overline{m}_o
\end{equation}
We then want to solve 
\begin{equation}
0 = \mathbf{m} - \mathbf{W}\mathbf{m}
\end{equation}
in a least-squares sense.  Expanding $\mathbf{m}$ in terms of $\mathbf{\minst}$, and rearranging, we obtain
\begin{equation}
(1 - \mathbf{W}) \mathbf{A} \mathbf{p} = (\mathbf{W} - 1) \mathbf{\minst}
\end{equation}
We solve this in a least-squares sense using a simple diagonal covariance matrix $\mathbf{C}$, with diagonal elements equal to the photometric variance in measurement $i$.  We impose an error floor of 0.01 mags on the photometric variances, to prevent a few bright stars from dominating the fit.  Letting $\mathbf{A^\prime} = (1 - \mathbf{W})\mathbf{A}$ and $\mathbf{b} = (\mathbf{W} - 1)\mathbf{\minst}$, this is an ordinary linear least squares problem with solution
\begin{equation}
\mathbf{p} = (\mathbf{A^{\prime\intercal}} \mathbf{C}^{-1} \mathbf{A^\prime})^{-1}\mathbf{A^\prime}\mathbf{C}^{-1}\mathbf{b}
\end{equation}
We solve this equation directly to find $\mathbf{p}$ and hence $Z_r$, the relative photometric calibration of the survey.  

The structure of $\mathbf{A^\prime}$ is illustrated in more detail in \citet{Padmanabhan:2008}.  We mention one appealing feature of the structure of the problem, however: if $\mathbf{A^\prime}$, $\mathbf{W}$, and $\mathbf{\minst}$ are written as sums of $\mathbf{A^\prime}_i$, $\mathbf{W}_i$, and $\mathbf{m}_{\mathrm{inst},i}$, where the terms in these sums contain only rows corresponding to observations of object $i$ and are otherwise $0$, then likewise $\mathbf{A^{\prime\intercal}} \mathbf{C}^{-1} \mathbf{A^\prime}$ and $\mathbf{A^\prime}\mathbf{C}^{-1}\mathbf{b}$ split into sums of terms involving only observations of a single object each.  This simplifies the computation of the matrices $\mathbf{A^{\prime\intercal}} \mathbf{C}^{-1} \mathbf{A^\prime}$ and $\mathbf{A^\prime}\mathbf{C}^{-1}\mathbf{b}$ and allows the terms contributing to them to be computed in parallel without ever requiring the matrix $\mathbf{A^\prime}$ to be built.  This is critically important because $\mathbf{A^\prime}$ has size $n_\mathrm{obs}\times n_\mathrm{par}$, which for a survey of a billion observations and a model containing thousands of parameters contains trillions of elements.  Because of the intrinsic parallelism of the problem, only $\mathbf{A^{\prime\intercal}} \mathbf{C}^{-1} \mathbf{A^\prime}$ needs to be built, which is of manageable size.  Likewise, the parallelism means that only the observations of a single object, and not the observations of the entire survey, need to be simultaneously read into memory.

The matrix $\mathbf{A^{\prime\intercal}} \mathbf{C}^{-1} \mathbf{A^\prime}$, computed as described above, is necessarily singular.  The singularity occurs because inevitably, for any solution $Z_r$, the solution $Z_r+c$ is equally good for any $c$.  The constant $c$ is degenerate with the absolute calibration of the survey.  Depending on the model for $Z_r$, other singularities may exist.  Accordingly, we perform a singular value decomposition of $\mathbf{A^{\prime\intercal}} \mathbf{C}^{-1} \mathbf{A^\prime}$.  Eigenvectors of $\mathbf{A^{\prime\intercal}} \mathbf{C}^{-1} \mathbf{A^\prime}$ below a certain threshhold are fixed using priors, as described in \citet{Padmanabhan:2008}.  The singularities affecting the PS1 calibration are described in \textsection \ref{subsec:algops1}.

This method correctly derives best-fit parameters $\mathbf{p}$ describing the relative photometric calibration.  The above discussion has described the photometric calibration algorithm in general terms.  We now describe the details of the calibration as applied to the PS1 survey.

\subsection{Details for \PS}
\label{subsec:algops1}

The most important question to be answered in applying the photometric calibration algorithm to a particular survey is the choice of model for $Z_r$.  The simplest reasonable model uses $Z_r = a_n - k_n x$, where $a_n$ describes the throughput of the optics, filters, and detector on a night $n$, and $k_n$ describes the transparency of the atmosphere on that night, as described in \textsection \ref{subsec:problem}.  We need to refine this model slightly for PS1.

In wide-field surveys, frequently the system response depends on the position of the star in the focal plane, for example, because of the need for an illumination correction \citep{Hogg:2001}.  The raw \PS\ images are already corrected for non-uniform detector throughput across the field of view according to a static flat field derived from the combination of dome flats and stellar photometry.  We nevertheless solve for a new flat field using the wealth of data taken since the beginning of the survey.  Comparison between SDSS and PS1 data indicated that the flat field changed abruptly three times during the survey \citep{Finkbeiner:2012}.  We accordingly fit for separate flat fields $f_{i,j}$ for the four different ``seasons'' to account for this behavior, so that $Z = a_n - k_n x + f_{i,j}$.  Here $i$ indexes over locations in the focal plane, which we take to be the four quadrants of each of the 60 CCDs in the PS1 focal plane, and $j$ indexes over the four seasons (Table \ref{tab:flatseasons}).

\begin{deluxetable}{ccc}
\tablecaption{Flat Field Seasons}
\tablehead{
\colhead{Season} & \colhead{Begin Date} & \colhead{End Date}
}
\startdata
Season 1 & --- & 10 April 2010 \\
Season 2 & 10 April 2010 & 11 May 2010 \\
Season 3 & 11 May 2010 & 11 April 2011 \\
Season 4 & 11 April 2011 & --- \\
\enddata
\tablecomments{
\label{tab:flatseasons}
The dates marking the boundaries of the different time periods for which independent flat fields are used in the photometric model, from \citet{Finkbeiner:2012}.  Season 1 includes all data taken before 10 April 2010, while season 4 includes all data taken after 10 April 2010.
}
\end{deluxetable}

Analysis of $Z$ derived for images taken as part of the medium-deep survey revealed that the amount of flux registered by the PS1 system systematically varied depending on the image quality of the individual images, in the sense that flux is lost for images with a very small or, especially, very large FWHM.  To account for this variation, we include in our model for $Z$ a quadratic $w$ in FWHM, leaving us with the model
\begin{equation}
\label{eq:zp}
Z = a_n - k_n x + f_{i,j} + w(F)
\end{equation}
where $F$ is the FWHM of the image in which the observation was made.  This model is summarized in Table \ref{tab:photomodel}, and is the final model we adopt to derive the \PS\ zero points.

\begin{deluxetable}{ccc}
\tablewidth{\columnwidth}
\tablecaption{Parameters of the Photometric Model}
\tablehead{
\colhead{Parameter} & \colhead{Number} & \colhead{Note}
}
\startdata
$a$ & $\sim 200$ & system (nightly) \\
$k$ & $\sim 200$ & atmosphere (nightly) \\
$f$ & $4\times60\times4$ & illumination correction \\
$w$ & 2 & FWHM correction (quadratic) 
\enddata
\tablecomments{
\label{tab:photomodel}
The parameters of the photometric model used in this work.  The calibration is performed independently in each of the 5 \PS\ filters.  Observations have been performed on about 200 nights in each filter, though the exact number ranges from 293 in \zps\ to 190 in \yps.  The illumination correction has one parameter describing each of the four quadrants of the 60 PS1 CCDs, over four time periods.  The constant term of the quadratic in $w$ is not fit, because it is completely degenerate with $a$.
}
\end{deluxetable}

We have experimented with fitting only a single $k$ term for the entire survey, rather than fitting one $k$ term for each night.  However, the best fit values of $k$ can vary from night to night, by about 0.05 mags/airmass (\textsection \ref{subsec:atmospherestability}).  That said, the median standard deviation in airmass $x$ for all observations on a given night is only about 0.1, leading to an induced uncertainty of about 5 mmag from ignoring the variation in $k$.  Nevertheless, because the extreme edges in declination of the survey, the north celestial pole and declination $-30\degree$, must be observed at relatively high airmass, neglecting the variation in $k$ leads to errors in the photometric calibration of the survey on large angular scales.  For this reason, we fit a $k$ term for each night of the survey.

Other models for $Z$ could be adopted.  In principle we have enough information to fit a zero point for every \PS\ image independently, though this would greatly diminish the stability of the solution on large angular scales.  However, the \PS\ system has proven to be remarkably stable over the course of a night (\textsection \ref{subsec:nightlystability}), removing the need for a more finely grained calibration, and leading us to adopt a simple model fitting only two parameters per night.

For the photometric calibration we use only observations taken at times when we believe the night to be photometric.  We define ``photometric'' here to mean that the simple model for $Z$ described above is satisfied on that night to within about 20 mmag.  In cloudy weather the $Z$ behave erratically; when we find evidence for clouds greater than 20 mmag in $Z$, we manually flag that portion of the night as non-photometric.  We flag about 25\% of the images taken as non-photometric.  We are able to recognize exposures as taken in non-photometric conditions only when they are discrepant with other overlapping \PS\ exposures, or when they overlap the SDSS.  As the \PS\ survey continues, our ability to flag non-photometric exposures will improve, though the analysis of \textsection \ref{sec:results} indicates that our performance is already good.

We wish to use only secure observations of typical stars in the calibration, to avoid any bias in the calibration from anomalous measurements.  We therefore use only objects for which at least one detection had estimated uncertainty less than 30 mmag.  We also exclude any objects for which any measurement of that object had an instrumental magnitude less than $-14.25$, to avoid any detector saturation effects.  We finally exclude any detections on images with FWHM greater than four arcseconds.

We use techniques for measuring \minst\ that assume that the object being measured is a point source (i.e., PSF mags).  Accordingly, we wish to include only point sources in the calibration, excluding the galaxies.  For this purpose, we exclude any object for which more than 25\% of the detections of that object have PSF magnitude minus aperture magnitude greater than $0.1$ mags.  We also exclude any objects for which more than 10\% of the detections of that object have $m > c$, with $c$ equal to 19 mag in \gps\ and \rps, 18.75 in \ips, and 18 in \zps\ and \yps.  These correspond approximately to the magnitudes at which the SDSS finds that the number density of galaxies exceeds the number density of stars, and so these cuts further reduce the galaxy contamination in our analysis.  Our tests indicate that resulting selection of stars is very clean, and varying the star-galaxy separation negligibly affects our results.

We iterate the photometric calibration several times and clip discrepant observations and images on each iteration, reducing the clipping threshhold by a factor of two on each iteration until reaching $3\sigma$, as described below.  This reduces the sensitivity of the algorithm to outliers.  In clipping detections, we compute $\Delta = m-\overline{m}$ for each detection.  For each image $i$, we find the mean $\mu_\Delta$ and standard deviation $\sigma_{\Delta}$ of $\Delta$ on that image.  An image is clipped if $\mu_\Delta$ or $\sigma_{\Delta}$ is inconsistent with their respective distributions for all images at the clipping threshhold.  About 3\% of images are clipped in this process; most periods of non-photometric conditions have already been flagged by hand.  A detection on an image is clipped if $|\Delta/\sigma|$ is greater than the clipping threshhold for that detection, where $\sigma$ is the sum in quadrature of the photometric uncertainty for that detection and $\sigma_{\Delta}$ for that image.  About 4\% of detections are clipped.  These include detections with problematic photometry and variable stars.

We modified the algorithm of \citet{Padmanabhan:2008} slightly to incorporate including an external source of photometry in the calibration.  When enabled, if external photometry for an object is available, we set $\overline{m}$ for that object to be that given by the external photometry, so that for these objects the best fit solution simply minimizes the difference between the external and internal photometry.  We can then add, for instance, all of the photometry from the SDSS matching PS1 objects into the photometric calibration.  We have experimented with both including and excluding the SDSS photometry.  In the final photometric calibration we currently use for PS1, we include the SDSS photometry, in order to improve our ability to detect non-photometric conditions.  The derived zero points vary only slightly depending on whether or not the SDSS is included as an external reference in the calibration, by about 5 mmag rms \textsection \ref{subsec:sdss}.

We impose Gaussian priors on the parameters of the photometric model when the photometry poorly constrains them, as described in Table \ref{tab:priors}.  We take the set of poorly constrained eigenvectors of the solution and find the best fit parameters for those eigenvectors such that the priors are satisfied, as described in \citet{Padmanabhan:2008}.  The priors serve three primary functions.  The first is to set the absolute calibration of the survey; the second is to resolve the degeneracy between the $a$ and $f$ terms; and the third is to resolve the degeneracy between the $a$ and $k$ terms on nights when the range of airmasses probed is small.

We impose a prior for the absolute calibration of the survey based on the work of \citet{JTphoto}, which uses photometry from standard stars and the Hubble Space Telescope to determine the absolute calibration of \PS.  We choose a prior for the $a$ terms to agree with the results of that work.  When using the SDSS as a reference, we impose this prior by using the color transformations of \citet{JTphoto} to transform the SDSS magnitudes into the \PS\ bandpasses, and then the absolute calibration is fixed directly by the photometric solution by reference to the color-corrected SDSS.

The $a$ and $f$ terms in the photometric model have an exact degeneracy, in that $Z$ is unchanged if all of the $a$ are increased and $f$ is decreased.  This degeneracy is removed by enforcing a prior on all of the $f$ to be zero, with an uncertainty of 20 mmag.

The $a$ and $k$ terms are degenerate if the range of airmass probed on a night is small.  This degeneracy is obviously exact on a night consisting of a single image.  Then $kx$ is just a constant for the night, and so any change in $a$ can be canceled with a change in $k$.  These degeneracies are removed by setting a prior on $k$ to be equal to the value of $k$ on a typical night.

\begin{deluxetable}{ccc|ccc}
\tablewidth{\columnwidth}
\tablecaption{Priors}
\tablehead{
\colhead{Parameter} & \colhead{Prior} & \colhead{$\sigma$} & \colhead{Parameter} & \colhead{Prior} & \colhead{$\sigma$}
}
\startdata
$a_g$ & 24.408 & 1 & $k_g$ & 0.147 & 0.05 \\
$a_r$ & 24.679 & 1 & $k_r$ & 0.085 & 0.03 \\
$a_i$ & 24.556 & 1 & $k_i$ & 0.044 & 0.02 \\
$a_z$ & 24.218 & 1 & $k_z$ & 0.033 & 0.02 \\
$a_y$ & 23.237 & 1 & $k_y$ & 0.073 & 0.03 \\
$f$ & 0 & 0.02 & $w$ & 0 & 0.1
\enddata
\tablecomments{
\label{tab:priors}
The Gaussian priors for the parameters in the photometric model.  The priors on $w$ are irrelevant as $w$ is always well constrained by the data.  The priors on $f$ serve only to break the degeneracy between $a$ and $f$.  The priors on $a$ set the absolute calibration of the survey, and are tuned to match the SDSS with the color transformations of \citet{JTphoto}.  The priors on $k$ constrain $k$ on nights when the range of airmasses probed by the survey is small, causing $k$ and $a$ to be degenerate.
}
\end{deluxetable}

After we obtain an iterated photometric solution, we perform a final adjustment to the derived $Z$ for each image.  We can robustly compute $\mu_{\Delta,i}$ for each image $i$ in the survey.  Ordinarily we use $\mu_{\Delta,i}$ to clip images that are discrepant from the rest of the photometric solution.  Having obtained a photometric solution, however, we use the $\mu_{\Delta,i}$ to improve the calibration of the survey by adjusting the photometric solution $Z_i$ for image $i$ by $\mu_{\Delta,i}$.  This induces no changes in the photometric calibration on large spatial scales, but does clean up the light curves of bright objects slightly.  The standard deviation of $\mu_{\Delta,i}$ is about 5 mmag (\textsection \ref{sec:results}).

\section{Results}
\label{sec:results}

We compute the PS1 photometric calibration using about one billion observations in each filter, solving for about a thousand parameters for each filter.  These parameters describe the system zeropoint $a$, atmospheric transparency $k$, and flat field, as well as an image quality correction as described by Equation \ref{eq:zp}.  We perform the computation in parallel over the available cores in our system, using the inherent parallelization described in \textsection \ref{subsec:algorithm}.  The computation time is dominated by reading data from disk for processing, and takes about two days.  The parallelization of the computation and the database operations more generally were greatly simplified by the Large Survey Database software \citep{Juric:2012}.

We find that the results of the photometric calibration account for variations in the mean zero point per exposure of the \PS\ system to better than 10 mmag rms, albeit with some areas of worse calibration.  We demonstrate this accuracy, checking the internal consistency of the solution (\textsection \ref{subsec:internal}), the consistency with the SDSS (\textsection \ref{subsec:sdss}), and the consistency of the colors of stars in different parts of the sky (\textsection \ref{subsec:colors}).

\subsection{Internal Consistency}
\label{subsec:internal}

We first test the internal consistency of the photometric calibration in three ways: by using simulated data, by examining the residuals from our photometric model over the sky, and by comparing with an alternative method for calibrating the MD fields.

Tests with simulated data indicate that errors due to the statistical uncertainty of the observations are negligible.  In these simulations, which do not include clouds, we take the actual set of observations of all of the stars used in this analysis.  For each star, we then declare its true magnitude for the purposes of the simulation to be equal to the mean of all measurements of that star.  We then convert these true magnitudes to instrumental magnitudes, using our adopted model for the zero points, with parameters drawn at random from the distributions given in Table~\ref{tab:priors}.  Noise is then added to these simulated measurements consistent with estimates from the actual \PS\ data, to produce a set of simulated \PS\ observations.

These simulated observations are then photometrically calibrated, and the recovered zero points for each observation are compared with the true zero points.  The standard deviation in the difference between the recovered magnitudes and true magnitudes is less than 0.2 mmag when using a simulated external reference catalog designed to match the SDSS.  If this simulated external reference catalog is not used, the standard deviation remains less than 0.2 mmag, though a small number of observations have zero points off by as much as 6 mmag in the \yps\ band.

These simulations do not include systematic errors caused by deviations in the system throughput from our model (for instance, clouds).  However, they verify that the data have the signal-to-noise necessary to constrain the model, and that the algorithm accurately recovers the photometric model parameters when given good data.

\begin{figure*}[tbh]
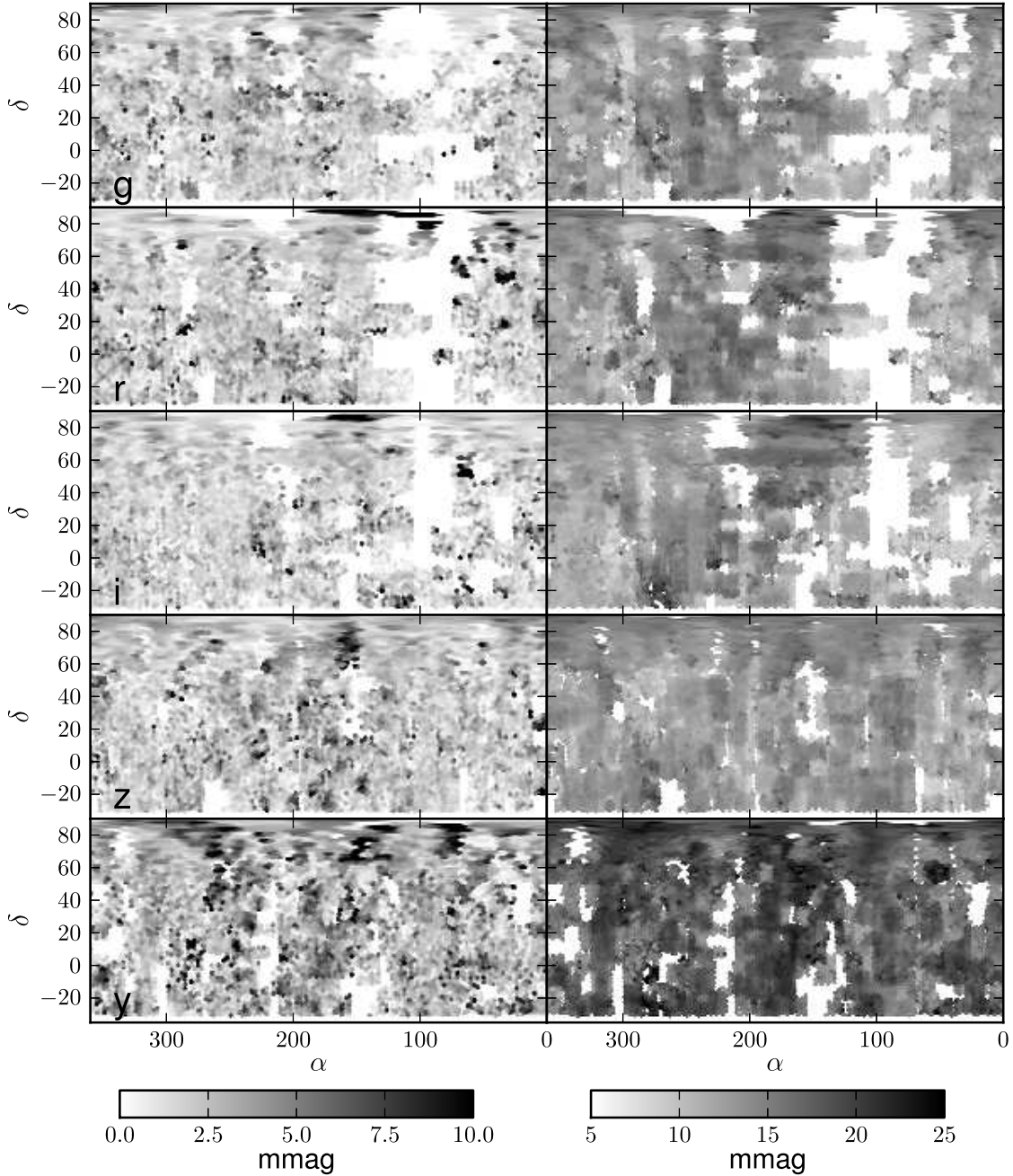

\dfplot{imstddevmap.eps}
\figcaption{
\label{fig:imstddevmap}
Maps of $|\mu_\Delta|$ (left panels) and $\sigma_\Delta$ (right panels) in $griz\yps$.  The x-axes give right ascension and the y-axes give declination, both in degrees.  These give the consistency of the zero points and the scatter in the photometry over the sky.  White to black is 0--10 mmag for the left panels and 5--25 mmag for the right panels.  The Galactic center is barely visible as a region of increased $\sigma_\Delta$ near (266\degree, $-29\degree$).  The $\sigma_\Delta$ are substantially larger in the \yps\ band than in the other bands.
}
\end{figure*}

We perform an internal test of the size of the deviations from our photometric model using $\Delta = m - \overline{m}$.  For each exposure, we compute $\mu_\Delta$ and $\sigma_\Delta$, the mean and standard deviation of $\Delta$ for all observations on a single exposure.  When an individual exposure has a zero point inconsistent with our photometric model, $\mu_\Delta$ will depart from zero.  Exposures with highly variable point spread functions or other problems with the photometry will have large $\sigma_\Delta$.  The typical size of $\sigma_\Delta$ will depend on the brightness of the stars used in the calibration; here we are interested only in verifying that the $\sigma_\Delta$ are homogeneous over the survey.  Figure \ref{fig:imstddevmap} shows a map of $|\mu_\Delta|$ and $\sigma_\Delta$ in each of the PS1 bands, using a PS1 calibration that does not include the SDSS as an external reference.  The maps of $|\mu_\Delta|$ are fairly uniform over the sky; a few isolated areas are slightly problematic, but no large scale trends are obvious and the Galactic plane makes no appearance.  The maps of $\sigma_\Delta$ are somewhat patchier and the Galactic center appears as an area of only slightly increased $\sigma_\Delta$, a testament to the performance of the IPP in crowded fields.  The $\sigma_\Delta$ have the tendency to be larger in areas with more observations, despite correction of $\sigma_\Delta$ by $\sqrt{N/(N-1)}$, where $N$ is the number of observations of the sky.  This is likely because exposures taken as part of TTI pairs are especially photometrically consistent; the two exposures are taken only fifteen minutes apart with a common position angle, and so each detection lands on nearly the same pixels in the focal plane, and the point spread function has had little time to change.

\begin{figure*}[tbh]
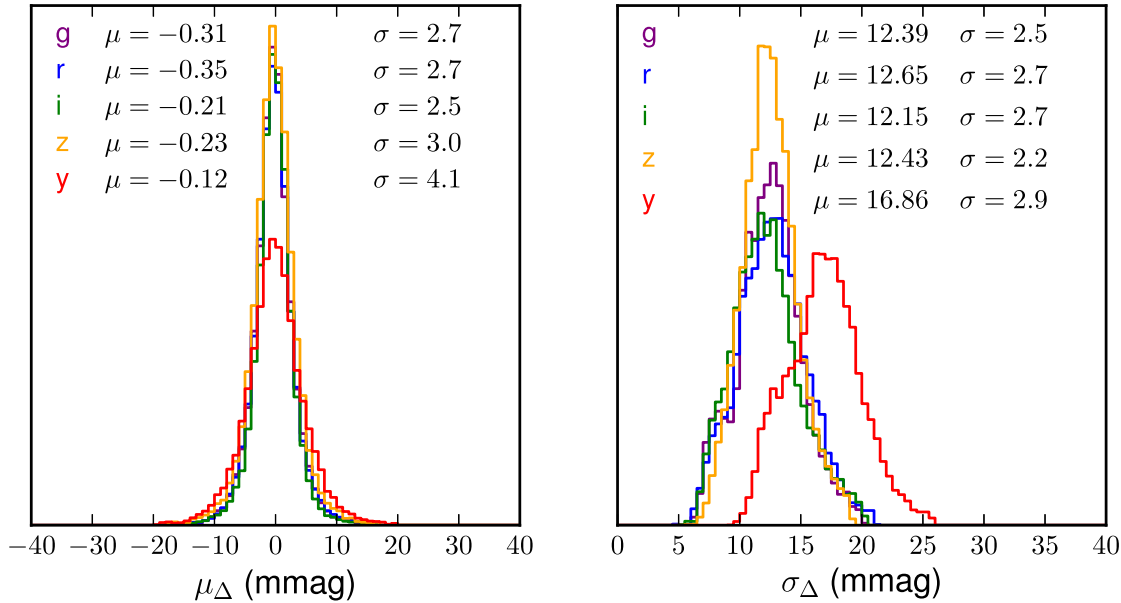

\dfplot{mnsdhists.eps}
\figcaption{
\label{fig:imstddevhist}
Histograms of $\mu_\Delta$ (left) and $\sigma_\Delta$ (right) in $griz\yps$.  The mean $\mu$ and standard deviation $\sigma$ of $\mu_\Delta$ and $\sigma_\Delta$ are labeled for each filter.  The scatter in $\mu_\Delta$ represents a lower bound on the uncertainty in our photometric calibration (\textsection \ref{subsec:internal}).  The $\sigma_\Delta$ are quite similar among the various bands, except in \yps, which has 50\% more scatter than the other bands.
}
\end{figure*}

Figure \ref{fig:imstddevhist} shows the same information as in Figure \ref{fig:imstddevmap} as a histogram, giving the distribution of $\mu_\Delta$ and $\sigma_\Delta$.  Unsurprisingly, $\overline{\mu_\Delta}$ is near zero.  The scatter in $\mu_\Delta$ is only about 3 mmag.  This is a lower bound on the actual uncertainty in our photometric calibration.  The $\sigma_\Delta$ are about 12 mmag, except in \yps, where the photometric scatter is about 17 mmag.

The scatter in $\mu_\Delta$ will be lower than the true uncertainty in the calibration because many stars have only been observed a few times, and often as part of correlated TTI pairs.  We can eliminate this problem by limiting the exposures used to compute $\mu_\Delta$ to exposures on MD fields.  In this case, the scatter in $\mu_\Delta$ goes up to 5 mmag, while $\sigma_\Delta$ remains about the same.  This scatter in $\mu_\Delta$ is more realistic, and represents the accuracy of the photometric calibration that we could expect after covering the sky many times.  However, sparsely covered portions of the sky may have photometric calibration errors much larger than found in these well-covered MD fields.

As a final internal consistency check, we compare the zero points generated by the photometric calibration algorithm with the zero points for the MD fields of \citet{Finkbeiner:2012}.  In that work, a very loose photometric model is adopted, which allows completely independent zero points for each exposure and a separate flat field for each night.  We find that the zero points of this work agree with those of \citet{Finkbeiner:2012} to about 5 mmag, consistent with our expectations from our internal tests of the model residuals on the MD fields.

\subsection{Consistency with the SDSS}
\label{subsec:sdss}

We can also verify the results of the photometric calibration by comparing our zero points with those we would derive by forcing the photometry to match an external reference as closely as possible.  The SDSS has observed about one third of the sky, about half of the area that \PS\ has observed.  We find zero points $Z_\mathrm{SDSS}$  for individual PS1 exposures that overlap the SDSS.   We compare these zero points with the zero points obtained from the photometric calibration algorithm, performed without using the SDSS as a reference.

We compute $Z_\mathrm{SDSS}$ by transforming the SDSS magnitudes of stars onto the \PS\ system using the color transformations of \citet{JTphoto}.  For each PS1 exposure, $Z_\mathrm{SDSS} = \langle m_\mathrm{SDSS} - m_\mathrm{inst} \rangle$.  We use only SDSS stars with $m_\mathrm{SDSS} < 18$ in the computation of $Z_\mathrm{SDSS}$.    

\begin{figure*}[tbh]
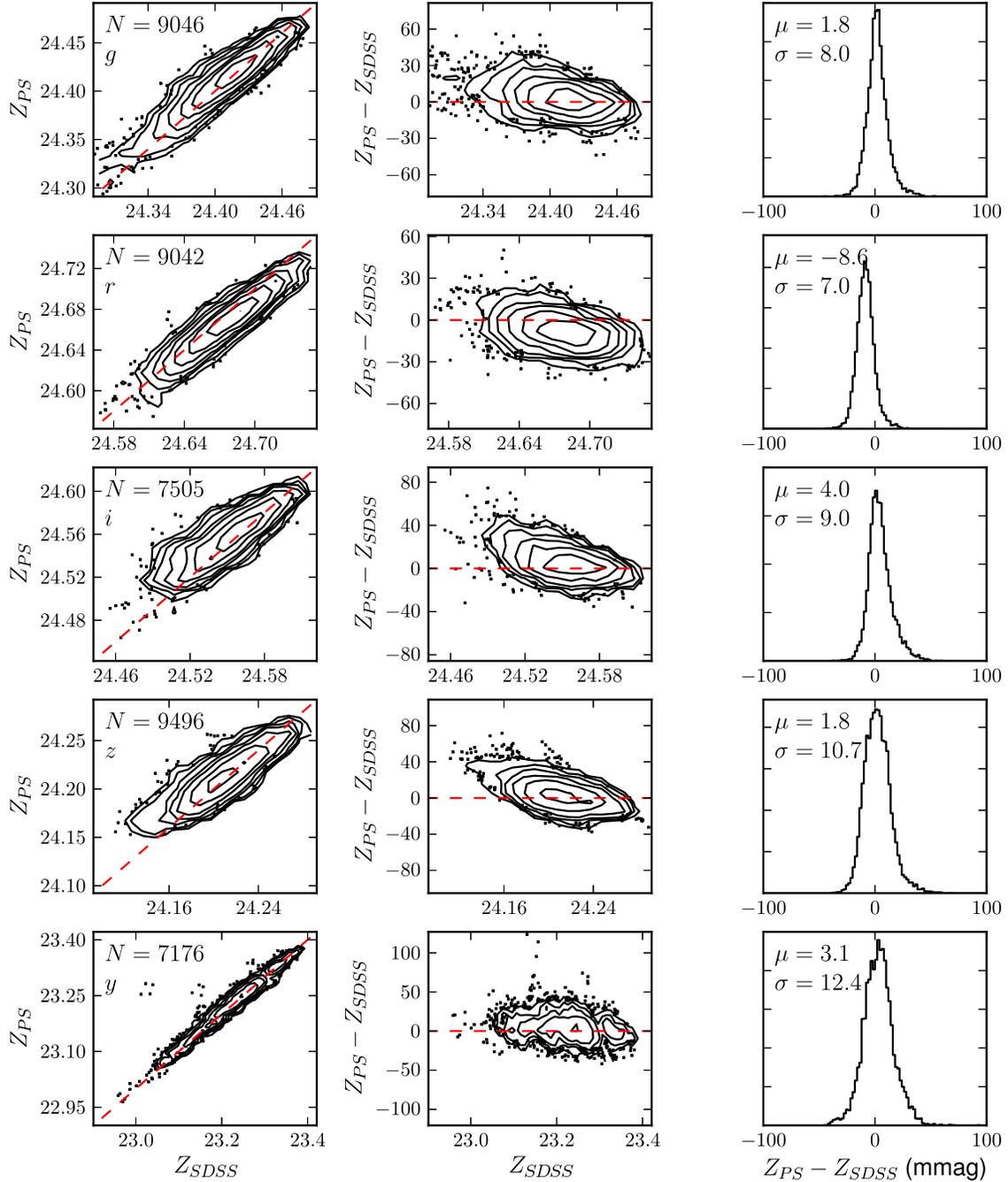

\dfplot{sdsscompplotsqt_noref.eps}
\figcaption{
\label{fig:sdsscompplots}
Comparisons between the zero points of this work and zero points derived relative to the SDSS, for the filters $griz\yps$ (rows).  The left column gives the zero points of this work relative to the zero points derived from the SDSS, both in magnitudes.  It also gives the number $N$ of images used in the comparison.  The middle column gives the difference in mmag between the two zero points with the SDSS zero points, in magnitudes.  The third column gives a histogram of the differences in mmag, along with their mean $\mu$ and standard deviation $\sigma$.  Unrecognized non-photometric conditions lead to points with low SDSS-derived zero points and large positive differences between SDSS- and PS-derived zero points.
}
\end{figure*}

The results of the comparison are shown in Figure \ref{fig:sdsscompplots}.  The internal \PS\ zero points agree with the SDSS-based zero points to about 10 mmag in all bands, ranging from 7 mmag in \rps\ to 13 mmag in \yps.  There are slight offsets in the mean zero point between the internal and SDSS zero points.  These means are determined by the absolute calibration of the survey, and are fixed by a prior; they provide no information about the relative calibration that is the focus of this work.  A small number of large photometric outliers do exist, and may be removed as the number of overlapping observations in the survey increases.

\begin{figure*}[tbh]
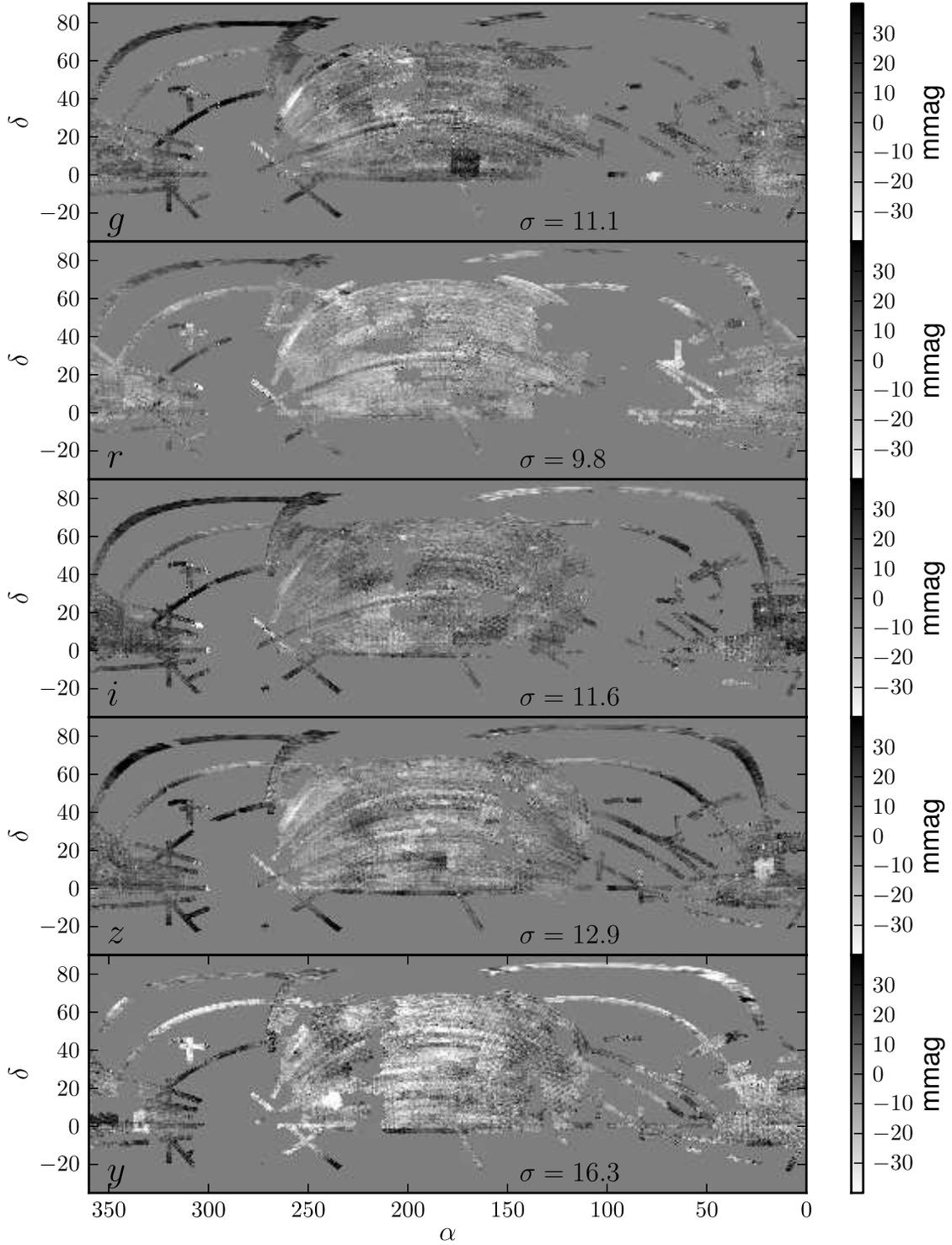

\dfplot{sdsscompmaps_qt.eps}
\figcaption{
\label{fig:sdsscompmaps}
Maps of the difference between the color-corrected SDSS magnitudes of stars and the internally-calibrated \PS\ magnitudes of the same stars in the filters $griz\yps$ (rows).  The x-axes give right ascension and the y-axes give declination.  The rms of the maps is about 10 mmag.  Narrow stripes in right ascension are symptomatic of problems with the SDSS photometric calibration, while rectangles in right ascension and declination indicate problems with the PS1 calibration.  The filter used for each map is indicated in the lower left, while the rms of the map is indicated in the lower right.
}
\end{figure*}

The spatial structure of the differences between the internal and SDSS-based zero points is of particular interest.  Figure \ref{fig:sdsscompmaps} shows maps of the mean difference between the calibrated PS1 magnitudes of objects used in the photometric calibration and their color-transformed SDSS magnitudes, in pixels $0.2\degree$ on a side.  The maps clearly show signs of errors in both the PS1 and SDSS photometric calibration, suggesting that a simultaneous PS1-SDSS calibration would be valuable.  The SDSS scan pattern is visible as narrow $3\degree$ stripes, approximately in right ascension, while the PS scan pattern is seen as rectangles in right ascension and declination.  The largest problems with the SDSS involve runs poorly connected to the main body of the SDSS; runs around $\alpha = 300\degree$ differ between PS1 and the SDSS by about 40 mmag.  The internal PS1 calibration shows clear $\sim 25$ mmag problems in \gps\ at (170\degree, 5\degree) and in \zps\ at (25\degree, 10\degree), to name a few, probably due to unrecognized cloudy weather.  Moreover, in \zps\ and \yps\ especially, parts of the maps are mottled at the $3\degree$ scale of the PS1 focal plane, indicating a problem with the PS1 photometry.  Despite these problems, the overall agreement between the two sets of measurements is remarkable; the rms of these maps is about 11, 10, 11, 12, and 16 mmag in $griz\yps$.  The results presented in Figure \ref{fig:sdsscompplots} are slightly better than these, because for that figure zero points were calculated for individual PS1 images, averaging over multiple SDSS runs and over the PS1 photometric nonuniformity in the focal plane.

\subsection{Consistency of Stellar Colors}
\label{subsec:colors}

We can additionally check the accuracy of the photometric calibration by testing the consistency of the stellar locus over the sky.  This technique has the advantage over comparison with the SDSS that it can be applied over the entire sky.  However, the presence of dust and intrinsic variations in stellar populations can cause the colors of the stellar locus to vary, limiting the effectiveness of this technique.

We measure the color of main-sequence turn-off stars over the sky to test the consistency of the stellar locus, following the technique of \citet{Schlafly:2010}.  This is one of a number of related techniques; we could alternatively have used the principal color analysis of \citet{Ivezic:2004} or the stellar locus regression of \citet{High:2009}.  Figure \ref{fig:btmaps} shows maps of the color of the blue tip of the stellar locus in the PS1 bands $griz\yps$.

\begin{figure*}[tbh]
\dfplot{btmaps_qt.eps}
\figcaption{
\label{fig:btmaps}
Maps of the color of the main-sequence turn-off (MSTO) stars observed by \PS.  The four rows give the colors $g-r$, $r-i$, $i-z$, and $z-y$.  The left column gives the observed MSTO color, while the right column gives the color corrected for dust according to \citet{Schlegel:1998} and \citet{Schlafly:2011}.  Clearly the dust accounts for most of the signal in these maps, but problems with the photometric calibration are also evident, for example, in \yps\ at (25\degree, $-25\degree$).  Blank areas have no \PS\ observations in photometric weather in one of the two bands making up the relevant color.  Black and white are $\pm 0.1$ mag of the median color of each map.
}
\end{figure*}

The dominant signal in Figure \ref{fig:btmaps} is clearly from the interstellar dust.  After removing the dust according to \citet{Schlegel:1998} and \citet{Schlafly:2011}, at high Galactic latitudes most of the signal comes from problems with the photometric calibration.  The most egregious example is in the \yps\ band at (25\degree, -25\degree), of 40 mmag.  Over most of the sky, calibration errors are consistent with expectations from the comparison with the SDSS, $\sim 15$ mmag.

\section{Discussion}
\label{sec:discussion}

We interpret these results in the context of the stability of the \PS\ system.  The system stability divides naturally into four different components: the photometric stability over a single night (\textsection \ref{subsec:nightlystability}), the stability of the detector and optics over the course of the survey (\textsection \ref{subsec:systemstability}), the stability of the atmosphere over the course of the survey (\textsection \ref{subsec:atmospherestability}), and the stability of the flat field over the survey (\textsection \ref{subsec:flatstability}).

\subsection{Nightly Photometric Stability of PS1}
\label{subsec:nightlystability}

\begin{figure*}[tbh]
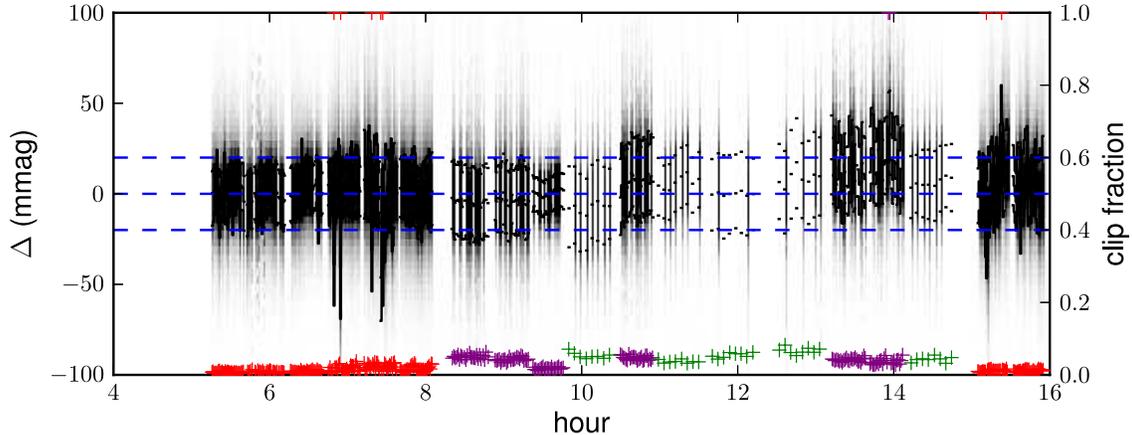

\dfplot{qa_night240.eps}
\figcaption{
\label{fig:qaplot}
The performance of the \PS\ photometry on 13 February 2010.  The figure gives the distribution of residuals $\Delta$, in mmag, of magnitudes of stars as observed on this night from the mean magnitudes of these stars, as a function of hour during the night.  Each column in the figure corresponds to an image.  The contours give the mean and $\pm 1\sigma$ of $\Delta$ for each image.  Crosses give the fraction of observations in each image clipped from the calibration, with 0\% corresponding to the bottom of the plot and 100\% the top of the plot.  The colors of the crosses give the band the image was taken in; the colors are the same as in Figure \ref{fig:zpstability}.  The mean of $\Delta$ for each image is small throughout the night, except for a 20 mmag deviation around hour 14.  This is one of the largest remaining photometric calibration errors in the survey.  See \textsection \ref{subsec:nightlystability} for details.
}
\end{figure*}

The photometric model we have adopted to calibrate the survey assumes that the throughput of the system and atmosphere do not vary substantially over the course of a night.  This assumption is occasionally violated, leading us to remove about 25\% of the nights on which survey data are taken.  The stability of the system on the remaining nights is excellent.  Figure \ref{fig:imstddevhist} showed that the typical model residuals are less than 5 mmag.  However, the nightly stability can be shown more explicity in a plot of model residual as a function of time of night.  Figure \ref{fig:qaplot} shows a variety of plots of data taken on February 13, 2010, during the first month of full science operation of the survey.  The first panel shows a density plot giving the distribution of $\Delta$ in mmag for each star in each image as a function of the hour in the night, with contours marking the mean and $\pm 1\sigma$ of $\Delta$ for each image.  Dashed blue lines mark $\pm 20$ mmag.  The crosses indicate the fraction of the detections clipped from each image in the photometric calibration, and their color indicates the band that each image was taken in, using the same colors as in Figure \ref{fig:zpstability}.  The photometric stability on this night is very good; the majority of images have $|\mu_\Delta| < 5$ mmag.  The \gps\ images taken around 14 hours, however, are discrepant at the 20 mmag level; this is the source of the clear \PS\ calibration error evident at (170\degree, 10\degree) in the $g$ band in Figure \ref{fig:sdsscompmaps}, and seems to be one of the largest calibration errors remaining in the data.

\subsection{System Stability}
\label{subsec:systemstability}

The scatter in the zero-point residuals from the simple nightly photometric model we adopt is about 5 mmag, showing that the system is stable over the course of a night.  However, this stability is not simply a nightly phenomenon; the system has been very stable over the whole course of the survey.

Ideally, we would measure the stability of the system as the scatter in the $a$-terms derived by the photometric calibration.  This procedure can overestimate the true scatter in the system throughput, however, because of the degeneracy between the $a$- and $k$-terms on nights when the range of airmass probed by the survey is small.  The median standard deviation of airmass in images taken in a single filter on a single night is only 0.1, rendering $a$ and $k$ substantially degenerate on most nights.  Accordingly, we instead test the stability of the system by looking at the scatter in the zero points $Z$ at $1.2$ airmasses, which is close to the average airmass of the survey.  We ignore the dependence of $Z$ on $w$ and $f$; the former could contribute slightly to the scatter in $Z$, but should play only a minor role, while the latter is required to have mean zero.  This combination of $a$ and $k$ is well-constrained, but includes variation both from $a$ and from $k x$.  The scatter from $k x$ at 1.2 airmasses is $1.2\sigma_k$.  We find in \textsection \ref{subsec:atmospherestability} that $\sigma_k$ is less than about 0.05, so the atmosphere should contribute less than 60 mmag scatter to the zero points.

\begin{figure}[tbh]
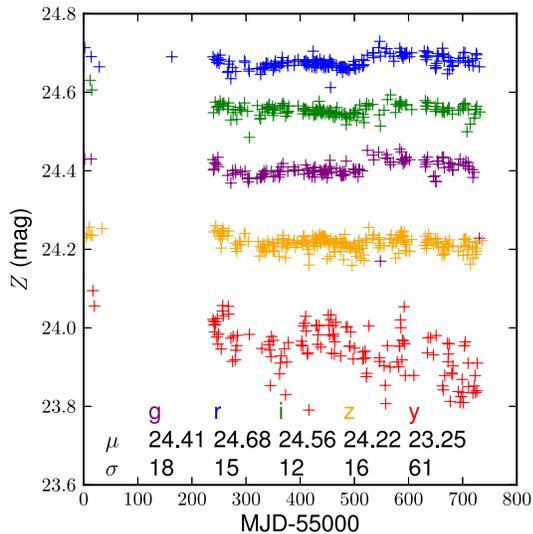

\dfplot{zpstability.eps}
\figcaption{
\label{fig:zpstability}
The system zero point $a-kx$ at airmass 1.2 derived by the photometric calibration for each night of the \PS\ survey.  The mean $\mu$ (in mag) and standard deviation $\sigma$ (in mmag) of the zero points is given for each band.  The \yps zero points have been offset by 0.7 mags for legibility.  The intrinsic system stability is better than 20 mmag in $griz$.  The \yps\ band is by far less stable than the other bands, with 61 mmag of scatter in zero point, presumably owing to the sensitivity of \yps\ to water in the atmosphere (see \textsection \ref{subsec:systemstability}).
}
\end{figure}

Figure \ref{fig:zpstability} shows the zero points $a-kx$ at airmass 1.2 for each photometric night of the survey in the 5 \PS\ bands.  The \yps\ bands have been offset by 0.7 mags for clarity.  The zero points are extraordinarily stable, except in \yps.  In the other bands, the intrinsic throughput of the \PS\ system has varied by less the 20 mmag rms.  A change in the system throughput near Modified Julian Day 55524 of 20 mmag is obvious in $griz$.  This stability is surprising given that we expect as much as 60 mmag scatter in zero point from the airmass term alone, and requires that the scatter in $k$ is actually less than 0.02 in $gri\zps$.

The \yps\ band has by far greater scatter in its nightly zero points than the other bands.  This is presumably the result of two factors, both owing to the presence of strong water vapor absorption bands that overlap \yps\ \citep{JTphoto}.  First, because the atmospheric absorption is not approximately constant over \yps, modeling the \yps\ zero point simply as linear in airmass is not appropriate.  Second, the depth of the absorption bands varies with the amount of precipitable water vapor in the air, leading to varying zero points.  An improved photometric calibration would then include more freedom in the \yps\ zero points, potentially incorporating additional information about water vapor in the air, as has been proposed in a number of works \citep{Stubbs:2007, Burke:2010, Blake:2011}, and adding color-airmass and color-water vapor terms to the photometric model.

The work of \citet{JTphoto} considers the dependence of zero point on airmass in more detail.  In particular, that work finds that adopting 
\begin{equation}
2.5\log{T_a} = -kx^{0.68}
\end{equation}
more accurately describes the relation between zero point and airmass in the \yps\ band.  However, we find that owing to the small range of airmass probed on a single night, adopting this relation alters the zero points derived in this work only negligibly ($\sim 1$ mmag).  Even so, we note that extrapolating the PS1 magnitudes of stars to the magnitudes that would be observed above the atmosphere requires taking the nonlinearity into account.  That problem is one of absolute calibration; in this work we consider only the relative calibration of the survey.

\subsection{Atmospheric Stability}
\label{subsec:atmospherestability}

The stability of the atmosphere over Haleakala places a fundamental limit on the stability of the \PS\ system.  The atmospheric transparency, however, can only be separated reliably from the system throughput when a wide range of airmass is probed over the course of a night.  On the typical night, however, the standard deviation in airmass of all observations in a single filter is less than 0.1.  In the \ips\ and \zps\ bands, this corresponds to less than 5 mmag of atmospheric extinction, comparable in size to the typical model residuals, making it impossible to reliably measure the airmass term $k$ in the photometric model for that night.

The problem is further complicated by the fact that the photometric calibration has access only to the differences in magnitudes of observations of the same stars.  Even when a night covers a wide range of airmass, if on another night the same stars were observed at the same airmasses, the calibration could reliably determine that those nights had the same $k$-term but not what that $k$-term was.

We use two techniques to test the variation in $k$ from night to night.  We limit ourselves to nights where $\sigma_x$, the standard deviation in airmass examined on that night, is greater than 0.1, and then look at the standard deviation in $k$ for those nights.  We determine $k$ from our photometric model, as well as using only zero points derived from the SDSS (\textsection \ref{subsec:sdss}).  

Measurements of $k$-terms derived from PS1-SDSS comparisons on nights with $\sigma_x > 0.1$ indicate that $\sigma_k$ is 0.03, 0.04, 0.03, 0.07, and 0.07 in $griz\yps$.  These results are sensitive to the limit on $\sigma_x$; taking $\sigma_x > 0.2$ results in $\sigma_k$ equal to 0.03, 0.03, 0.03, 0.04, 0.09, though only about ten nights of the survey have $\sigma_x$ that large.  Reducing $\sigma_x$ increases $\sigma_k$ in all bands, presumably because $\sigma_k$ becomes dominated by slight deviations from the photometric model.  These estimates of $\sigma_k$ all exceed our estimates for $\sigma_k$ based on the stability of the total throughput at 1.2 airmass (\textsection \ref{subsec:systemstability}), suggesting that the 20 mmag of scatter in zero points is dominated by atmospheric variations rather than variations in the PS1 system, and that the SDSS-based $\sigma_k$ are overestimates.

We can repeat this analysis using $k$-terms derived directly from the photometric calibration.  Because the PS1 survey region covers a larger range of declination than the SDSS covers, this allows more high-airmass observations to be included.  Using nights with $\sigma_x > 0.1$ indicates that $\sigma_k$ is $0.02$, $0.02$, $0.02$, $0.04$, and $0.05$ in $griz\yps$.  Using only nights with $\sigma_x > 0.2$ gives $\sigma_k$ equal to $0.02$, $0.02$, $0.02$, $0.02$, and $0.04$, which is consistent with our constraint from \textsection \ref{subsec:systemstability}.  If $\sigma_k$ in \yps\ is in fact about 0.04, then both the \PS\ system and atmosphere are less stable in \yps\ than in $gri\zps$.

\subsection{Detector Stability}
\label{subsec:flatstability}

The stability of the \PS\ detector can be tested independently from the atmosphere.  The raw PS1 images are flat-fielded with a single, static flat field derived from dome flats and stellar photometry taken early in the survey.  The photometric calibration derives four independent flat fields for different time periods in the survey (Table \ref{tab:flatseasons}).  The time variation in these flat fields tests the stability of the detector.  

\begin{figure}[tbh]
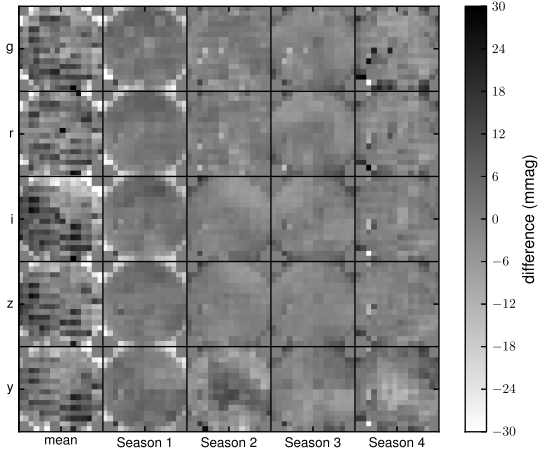

\dfplot{flatplot.eps}
\figcaption{
\label{fig:flats}
The flat fields, in mmag, derived by the photometric calibration, in the $griz\yps$ filters.  The first column gives the mean flat field for the survey in each filter, while the later columns give the difference between the flat field for that season (Table \ref{tab:flatseasons}) and the mean flat field.  The standard deviation in the mean flat is about 10 mmag, and the mean absolute residual of the difference between this mean flat and four flat seasons is less than 5 mmag, in all of the filters.
}
\end{figure}

Figure \ref{fig:flats} shows the mean flat field and the difference from the mean flat field for the four flat field seasons in $griz\yps$.  Each image consists of $16\times16$ pixels, describing the four quadrants of each of the 60 OTA CCDs composing the PS1 focal plane.  The mean absolute difference of the flat for each season from the mean flat field is less than 5 mmag in all of the filters.  Likewise, the standard deviation of the differences of the flat fields from the mean is less than 6 mmag in all of the filters.  The bright ring present in the flat fields from the first season indicates that the edge of the focal plane in the first season was too faint and required correction.  This may be due to problems with the photometry in images with poor image quality, which was especially problematic early in the survey and around the edge of the focal plane.

The mean flat has a scatter of about 10 mmag in all of the filters.  Some striping in the mean flat suggests that one side of the PS1 OTA CCDs has a $\sim 10$ mmag different throughput from the other side.  These variations should have been removed by the static PS1 flat field, yet are nevertheless robustly present in the photometric calibration and in comparison with the SDSS.  This may point to an anomaly occuring during the observations of the stars used to construct the static PS1 flat field.

\section{Conclusion}
\label{sec:conclusion}

We present the photometric calibration of the first 1.5 years of \PS\ survey data.  The per-image zero points we measure agree with those computed independently relative to the SDSS with rms scatter of 8.1, 7.1, 9.0, 11.5, and 12.7 mmag in $griz\yps$.  This indicates that both surveys have zero points accurate at that level, when averaged over the 3\degree\ field of view of PS1.  On several arcminute scales, photometric nonuniformities over the PS1 field of view and striping in the SDSS start to contribute, but the rms scatter remains only 11, 10, 11, 12, and 16 mmag.  We anticipate that as the PS1 survey continues, the accuracy of this calibration will improve as repeat observations reveal slight deviations from the simple photometric model we adopt.  Internal tests of the calibration indicate that we may be able to achieve overall image zero point accuracy as good as 5 mmag.

This accuracy renders the photometric calibration a minor source of systematic error in the PS1 photometry.  \citet{Finkbeiner:2012} has discovered a non-linearity in the PS1 photometry that can bias the photometry of faint sources by a few hundredths when image quality is poor.  Likewise, poor image quality in some PS1 images lead to systematics in the  PSF magnitudes that dominate the photometric calibration errors.

This photometric calibration demonstrates that 10 mmag calibration accuracy is possible using the survey strategies typical of major upcoming surveys, like the SkyMapper \citep{Keller:2007}, the Dark Energy Survey \citep[DES]{Flaugher:2005}, the Hyper Suprime-Cam Survey \citep[HSC]{Takada:2010}, and the Large Synoptic Survey Telescope \citep[LSST]{Tyson:2002}.  These surveys operate in a mode much like PS1, repeatedly imaging the sky one filter at a time.  This is in contrast to the SDSS, which operated in a drift-scanning mode and nearly-simultaneously imaged the sky in each of its filters.  We note, however, that the simultaneous five-color imaging of the SDSS leads to the SDSS having better color accuracy than magnitude accuracy, and that it makes detecting non-photometric weather easier.

The calibration also demonstrates the possibility of photometrically calibrating the \yps\ band, which includes a strong water vapor absorption feature at about 940 nm.  Despite the variability of this feature, we still achieve zero point accuracy of 13 mmag in this band.  This figure may be further improved by including the amount of precipitable water vapor in the air into the calibration.  SkyMapper, DES, HSC, and the LSST all also intend to observe in filters including this same feature.

The zero points achieved by this work show that the PS1 system is photometrically stable.  Zero points extrapolated to 1.2 airmasses every photometric night have rms scatter of less than 20 mmag in $gri\zps$, over the course of the survey.  Moreover, these 20 mmag of scatter are dominated by the variability of the atmosphere.  The stability of the PS1 optical system and detector is particularly impressive given the continuous improvements to the system over its first 1.5 years of operation, to reduce sky brightness, improve image quality, and defeat camera artifacts.

The level of calibration accuracy we have achieved will enable many PS1 survey goals.  The discovery of satellites of the Milky Way, cosmological investigations of supernovae and galaxy clustering, and studies of interstellar reddening all require accurate photometric calibration.  When the PS1 data become public, this calibration will provide another benefit to the community: a set of well-calibrated observations of stars covering most of the sky.  Together with the absolute calibration of the PS1 data as described in \citet{JTphoto}, this work provides an absolutely calibrated set of standard stars over the entire sky north of declination $-30\degree$, going much fainter than current data sets.

D.F. and E.S. acknowledge support of NASA grant NNX10AD69G for this research.  M.J. acknowledges support by NASA through Hubble Fellowship grant \#HF-51255.01-A awarded by the Space Telescope Science Institute, which is operated by the Association of Universities for Research in Astronomy, Inc., for NASA, under contract NAS 5-26555.  NFM acknowledges funding by Sonderforschungsbereich SFB 881 ``The Milky Way System'' (subproject A3) of the German Research Foundation (DFG)

The Pan-STARRS1 Survey has been made possible through contributions of the Institute for Astronomy, the University of Hawaii, the Pan-STARRS Project Office, the Max-Planck Society and its participating institutes, the Max Planck Institute for Astronomy, Heidelberg and the Max Planck Institute for Extraterrestrial Physics, Garching, The Johns Hopkins University, Durham University, the University of Edinburgh, Queen's University Belfast, the Harvard-Smithsonian Center for Astrophysics, and the Las Cumbres Observatory Global Telescope Network, Incorporated, the National Central University of Taiwan, and the National Aeronautics and Space Administration under Grant No. NNX08AR22G issued through the Planetary Science Division of the NASA Science Mission Directorate.

Funding for SDSS-III has been provided by the Alfred P. Sloan Foundation, the Participating Institutions, the National Science Foundation, and the U.S. Department of Energy Office of Science. The SDSS-III web site is http://www.sdss3.org/.

SDSS-III is managed by the Astrophysical Research Consortium for the Participating Institutions of the SDSS-III Collaboration including the University of Arizona, the Brazilian Participation Group, Brookhaven National Laboratory, University of Cambridge, University of Florida, the French Participation Group, the German Participation Group, the Instituto de Astrofisica de Canarias, the Michigan State/Notre Dame/JINA Participation Group, Johns Hopkins University, Lawrence Berkeley National Laboratory, Max Planck Institute for Astrophysics, New Mexico State University, New York University, Ohio State University, Pennsylvania State University, University of Portsmouth, Princeton University, the Spanish Participation Group, University of Tokyo, University of Utah, Vanderbilt University, University of Virginia, University of Washington, and Yale University.

\bibliographystyle{apsrmp}
\bibliography{ucal}

\end{document}